\documentclass[fleqn,usenatbib]{mnras}

\usepackage{newtxtext,newtxmath}

\usepackage[T1]{fontenc}

\DeclareRobustCommand{\VAN}[3]{#2}
\let\VANthebibliography\thebibliography
\def\thebibliography{\DeclareRobustCommand{\VAN}[3]{##3}\VANthebibliography}

\usepackage{amsmath}	
\usepackage{ae,aecompl}
\usepackage{comment}

\usepackage{graphicx,tabularx}	
\usepackage{booktabs}
\usepackage{color}

\usepackage{newtxtext,newtxmath}

\title[Conditional Diffusion Model for Astrophysical images]{Can Diffusion Model Conditionally Generate Astrophysical Images?}

\author[X. Zhao et al.]{
Xiaosheng Zhao,$^{1,2}$\thanks{E-mail: zhaoxs18@mails.tsinghua.edu.cn (XZ)} 
Yuan-Sen Ting,$^{3,4,5}$\thanks{E-mail: yuan-sen.ting@anu.edu.au (YT)}
Kangning Diao$^{1}$
and Yi Mao$^{1}$\thanks{E-mail: ymao@tsinghua.edu.cn (YM)}
\\
$^{1}$Department of Astronomy, Tsinghua University, Beijing 100084, China\\
$^{2}$CNRS \& Sorbonne Université, Institut d’Astrophysique de Paris (IAP), UMR 7095, 98 bis bd Arago, F-75014 Paris, France\\
$^{3}$Research School of Astronomy \& Astrophysics, Australian National University, Canberra, ACT 2611, Australia\\
$^{4}$School of Computing, Australian National University, Canberra, ACT 2601, Australia\\
$^{5}$Department of Astronomy, The Ohio State University, Columbus, OH 43210, USA
}

\date{26 October 2023}

\pubyear{2023}

\begin{document}
\label{firstpage}
\pagerange{\pageref{firstpage}--\pageref{lastpage}}
\maketitle

\begin{abstract}

Generative adversarial networks (GANs) are frequently utilized in astronomy to construct an emulator of numerical simulations. Nevertheless, training GANs can prove to be a precarious task, as they are prone to instability and often lead to mode collapse problems. Conversely, the diffusion model also has the ability to generate high-quality data without adversarial training. It has shown superiority over GANs with regard to several natural image datasets. In this study, we undertake a quantitative comparison between the denoising diffusion probabilistic model (DDPM) and StyleGAN2 (one of the most robust types of GANs) via a set of robust summary statistics from scattering transform. In particular, we utilize both models to generate the images of 21~cm brightness temperature mapping, as a case study, conditionally based on astrophysical parameters that govern the process of cosmic reionization. Using our new Fréchet Scattering Distance (FSD) as the evaluation metric to quantitatively compare the sample distribution between generative models and simulations, we demonstrate that DDPM outperforms StyleGAN2 on varied sizes of training sets. Through Fisher forecasts, we demonstrate that on our datasets, StyleGAN2 exhibits mode collapses in varied ways, while DDPM yields a more robust generation. We also explore the role of classifier-free guidance in DDPM and show the preference for a non-zero guidance scale only when the training data is limited. Our findings indicate that the diffusion model presents a promising alternative to GANs in the generation of accurate images. These images can subsequently provide reliable parameter constraints, particularly in the realm of astrophysics.

\end{abstract}


\section{Introduction} 
\label{sec:intro}

Deep generative models are neural network-based architectures that learn to generate data samples that resemble a given dataset. They find extensive employment across diverse domains, encompassing weather forecasting \citep{weather}, rocket aerodynamics \citep{YAN2019826}, subsurface fluid simulation for oil extraction \citep{JIN2020107273}, and natural language generation \citep{radford2018improving}. In the realm of astronomy, the exigency for generative models becomes pronounced in the realm of inference problems, where the exploration of an immensely vast parameter space necessitates modeling. The generative models hold the potential to expedite the modeling process while mitigating computational costs. Among the most widely used generative models, generative adversarial networks (GANs, \citealp{Goodfellow_2014}) excel in producing superior-quality images when compared with likelihood-based alternatives such as variational autoencoders and normalizing flows, and has been broadly exploited in astronomy \citep{2020MNRAS.496L..54M, 2021MNRAS.506..357Y, 2021ApJ...906L...1M}. The success of GANs comes from the adversarial training that drives the generator to continuously improve its ability to create realistic images by engaging in a competitive learning process with the discriminator. Adversarial training in GANs, while highly effective, introduces challenges in the form of unstable training processes and the occurrence of mode collapse \citep{2017arXiv170100160G, NEURIPS2018_0172d289}.
In more recent times, Diffusion models \citep{10.5555/3045118.3045358, NEURIPS2020_4c5bcfec, pmlr-v139-nichol21a, NEURIPS2021_49ad23d1, rombach2021highresolution, 2023arXiv230111093H, pmlr-v162-nichol22a, song2021scorebased, 2023arXiv230111093H} have eclipsed GANs in terms of image diversity and quality, without the need for adversarial training process. Inspired by non-equilibrium thermodynamics, the diffusion model defines a diffusion process by gradually adding random noise to the data and learns to reverse the diffusion process to sample desired data from the noise (denoising). In the application of image generation, the denoising process is conducted by progressively producing cleaner images corrupted by varying degrees of noise, employing cutting-edge vision models like the U-Net \citep{2015arXiv150504597R} and the vision transformer \citep{2020arXiv201011929D}. This sequential denoising process during training leads to better coverage of the data distribution and more stable gradients.

Recently, diffusion models have also been applied to the realm of astronomy \citep{2022MNRAS.511.1808S, 2022arXiv221104365K, 2022arXiv221103812A, 2022arXiv221112444M, 2023A&A...672A..51R, 2023arXiv230203046L, 2023arXiv230403788L, 2023arXiv230401670X}. However, there exists a dearth of quantitative comparisons regarding conditional image generation between diffusion models and GANs in the domain of astronomy. In this study, we delve into one typical diffusion model, namely the denoising diffusion probabilistic model (DDPM, \citealp{NEURIPS2020_4c5bcfec}), augmented with the latest functionalities for conditional image generation, and compare it with one of the most robust GAN models, namely StyleGAN2 which is widely applied for cosmological emulations \citep{2022arXiv220604594J,diao2023multifidelity} or super-resolution simulations \citep{doi:10.1073/pnas.2022038118, 2023arXiv230512222Z}. This comparison is conducted in the context of the images of 21~cm brightness temperature mapping \citep{Furlanetto2006}, which embody intricate physical origins and are easily accessible through public simulations. Specifically, our focus is to investigate whether DDPM can accurately generate 21 cm images conditional on astrophysical parameters compared with StyleGAN2.

To facilitate a quantitative comparison, we employ the scattering transform \citep{mallat2012group,6522407,allys2019rwst,cheng2020new,2021arXiv211201288C,2021arXiv210309247C,2021arXiv210411244S}, known for its capacity to extract robust non-Gaussian information from images. The scattering transform borrows the mathematical concepts in convolutional neural networks (CNNs), obviating the necessity of a training process. It has also found application in the 21 cm field (\citealt{2022MNRAS.513.1719G, 2023MNRAS.519.5288G}), showcasing its superiority over two-point statistics. Building upon the scattering transform, we formulate a single-value metric to assess the quality of image generation, investigate the influence of classifier-free guidance \citep{ho2021classifier} on conditional DDPM, and employ Fisher forecasts \citep{10.2307/2342435} to further explore the potential of parameter forecasts and the distribution coverage of images. Based on these investigations, we reach the conclusion about which approach, if any, is better suited for astrophysical image generation and for simulation-based inference (SBI) from astrophysical images.

The rest of this paper is organized as follows. We give a review of the two generative models in Section~\ref{sec:generative}, summarize the scattering transform and introduce the Fréchet Scattering Distance in Section~\ref{sec:Sta_Met}. We describe the data preparation in Section~\ref{sec:simu}, present the results in Section~\ref{sec: results}, lead the discussion in Section~\ref{sec: discuss}, and reach the conclusion in Section~\ref{sec:conclusion}. The details about the network architectures and hyperparameters are left in Appendix~\ref{sec: net}.

\section{Generative models}
\label{sec:generative}

\begin{figure*}
    \centering
    \includegraphics[width=\textwidth]{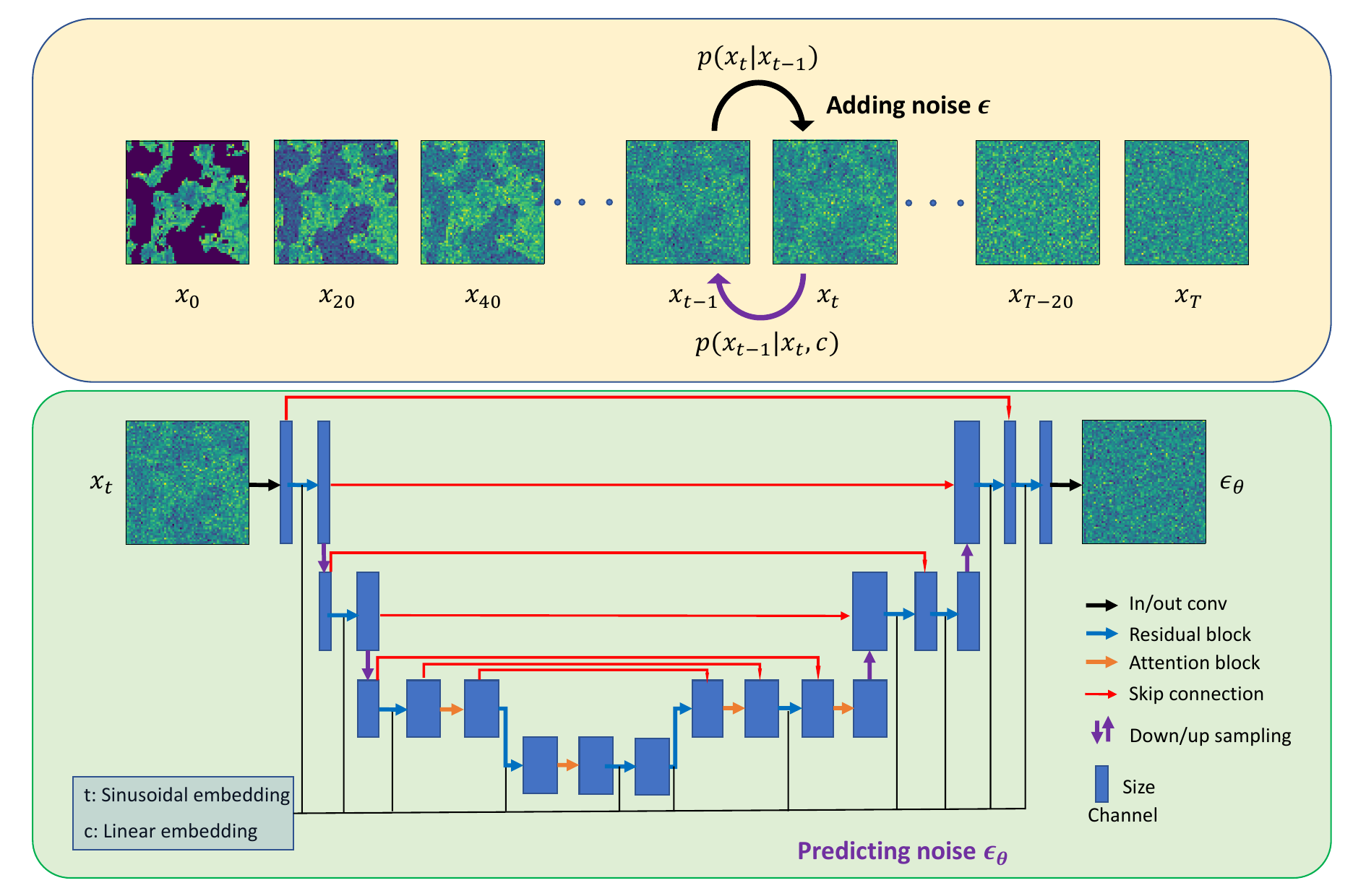}
    \caption{DDPM for image generations conditional on the astrophysical parameter $\mathbf{c}$. The upper panel shows the process of adding noise gradually and sampling images from the noise (denoising). The lower panel shows a simplified sketch of the U-Net model used to predict the noise in each denoising step. The sinusoidal embedding \citep{NIPS2017_3f5ee243} of the time $t$ and linear embedding of the conditional parameter $\mathbf{c}$ are added together and injected into each residual block. More details about the hyperparameter choice can be found in Appendix~\ref{sec: unet}.}
    \label{fig: diff_plot}
\end{figure*}

\subsection{Denoising diffusion probabilistic models}
Diffusion models are generative models that map the noise with a standard Gaussian distribution to the clean data with some target distribution iteratively. Starting from $
\mathbf{x}_{T} \sim \mathcal{N}(\mathbf{0}, \mathbf{I})$, $\mathbf{x}_{T-1}, \mathbf{x}_{T-2}, \mathbf{x}_{T-2}\dots\mathbf{x}_{0}$ are sampled given some neural networks outputs, where $T$ is the total denoising time steps, and $\mathbf{x}_{0}$ is the clean data that is in the form of images in this study.

The denoising diffusion probabilistic model (DDPM, \citealp{NEURIPS2020_4c5bcfec}) is one typical realization of diffusion models, where the noised versions of images are generated by $
\mathbf{x}_{t}\left(\mathbf{x}_{0}, \boldsymbol{\epsilon}\right)=\sqrt{\bar{\alpha}_{t}} \mathbf{x}_{0}+\sqrt{1-\bar{\alpha}_{t}} \boldsymbol{\epsilon}
$, for noise $
\boldsymbol{\epsilon} \sim \mathcal{N}(\mathbf{0}, \mathbf{I})
$ and $\bar{\alpha}_t$ some variance schedule parameter with the notation: $\alpha_{t}:=1-\beta_{t} \text { and } \bar{\alpha}_{t}:=\prod_{s=1}^{t} \alpha_{s}$. Here $\beta_{t}$ is another variance schedule parameter equally spaced between $\beta_{1}=10^{-4}$ and $\beta_{T}=0.02$. During the training stage, a neural network with parameter $\theta$ is used to predict the noise $\boldsymbol{\epsilon}_{\theta}$ and is optimized by the following loss function
\begin{equation}
L(\theta):=\mathbb{E}_{t, \mathbf{c}, \mathbf{x},  \epsilon}\left[\left\|\boldsymbol{\epsilon}-\boldsymbol{\epsilon}_{\theta}\left(\mathbf{x}_{t}, \mathbf{c}, t\right)\right\|^{2}\right]
\end{equation}
It is basically a mean square error where $\left\| \cdot \right\|$ is the $L_2$ norm and $\mathbb{E}_{t, \mathbf{c}, \mathbf{x}, \epsilon}\left[\cdot\right]$ is the expectation of the square error over the denoising time step $t$ which is drawn from a uniform distribution $\mathcal{U}[1, T]$. Here we choose $T=1000$. At each time step, the square error is parameterized by $\mathbf{c}, \mathbf{x}, \epsilon$ where $\mathbf{c}$ is the conditional information drawn together with the image $\mathbf{x}_{0}$ from the training dataset. 

During the sampling stage, denoised versions of images are generated by 
\begin{equation}
\mathbf{x}_{t-1}=\frac{1}{\sqrt{\alpha_{t}}}\left(\mathbf{x}_{t}-\frac{1-\alpha_{t}}{\sqrt{1-\bar{\alpha}_{t}}} \boldsymbol{\epsilon}_{\theta}\left(\mathbf{x}_{t}, \mathbf{c}, t\right)\right)+\sigma_{t} \mathbf{z}
\end{equation}
for the random noise and the final noised image $ \mathbf{z}, \mathbf{x}_{t} \sim \mathcal{N}(\mathbf{0}, \mathbf{I})$, $\sigma_{t}=\sqrt{\alpha_{t}}$ the further variance schedule parameter. For the noise prediction, we do not directly use the U-Net model in \citet{NEURIPS2020_4c5bcfec}, but build the U-Net model based on the improved version described in \citet{NEURIPS2021_49ad23d1}. The detailed hyper-parameter choice for the U-Net model can be found in Appendix~\ref{sec: net}. We embed our two parameters as the condition information and add them to the time embedding. The added tensor is used as the auxiliary input of the conditional U-Net. We show the architecture of our conditional diffusion model in Fig.~\ref{fig: diff_plot}.

\subsubsection{Classifier-free guidance}
Classifier-free guidance is introduced \citep{ho2021classifier} for text-to-image generation and also used in popular models like Stable Diffusion \citep{rombach2021highresolution} and Imagen \citep{saharia2022photorealistic}. It is shown to achieve a better trade-off between sample quality and diversity. During training time, the label is discarded randomly with a probability (0.28 in this work). So the networks learn both the unconditional and conditional generative models. During the sampling time, the noise prediction $\tilde{\boldsymbol{\epsilon}}_{\theta}$ is a combination of the unconditional prediction $\boldsymbol{\epsilon}_{\theta}\left(\mathbf{x}_{t}\empty\right)$ and conditional prediction $\boldsymbol{\epsilon}_{\theta}\left(\mathbf{x}_{t}|\mathbf{c}\right)$ with a guidance scale $s$
\begin{equation}
\tilde{\boldsymbol{\epsilon}}_{\theta}\left(\mathbf{x}_{t}|\mathbf{c}\right)=(1+s) \boldsymbol{\epsilon}_{\theta}\left(\mathbf{x}_{t}|\mathbf{c}\right)-s \boldsymbol{\epsilon}_{\theta}\left(\mathbf{x}_{t}\empty\right),
\label{eq: guidance}
\end{equation}
where $s=0$ leads to a standard conditional generative model, and $s=-1$ leads to a standard unconditional generative model. Note that the above formula can be rewritten by $\tilde{\boldsymbol{\epsilon}}_{\theta}\left(\mathbf{x}_{t}|\mathbf{c}\right)= \boldsymbol{\epsilon}_{\theta}\left(\mathbf{x}_{t}\empty\right)+s^\prime ( \boldsymbol{\epsilon}_{\theta}\left(\mathbf{x}_{t}|\mathbf{c}\right)-\boldsymbol{\epsilon}_{\theta}\left(\mathbf{x}_{t}\empty\right))$ \citep{pmlr-v162-nichol22a}, where $s^\prime = 1$ leads to a standard conditional generative model. From this form, the guidance scale can be interpreted by guiding the unconditional model to the direction where the output image meets the input condition \citep{dieleman2022guidance}. When $s^\prime > 1$, or equivalently, $s > 0$ in Equation~\ref{eq: guidance}, the diffusion models tend to conditionally generate higher-quality natural images, but at costs to image diversity.

\subsection{StyleGAN}
\label{sec: stylegan}
A GAN model is composed of two deep neural networks: a generator $G$ and a discriminator $D$. The GAN loss function used in our work is adversarial loss. A conditional version can be formulated as

\begin{equation}
    \mathcal{L}_{\rm adv} = D(G(\mathbf{z},\mathbf{c})|\mathbf{c}) - D(\mathbf{x}|\mathbf{c})\\
\end{equation}
The training objective is
\begin{equation}
    (G^*,D^*) = \arg \min\limits_{G} \max \limits_{D} \mathbb{E}_{\mathbf{z}\sim p(\mathbf{z}),\mathbf{x}\sim p(\mathbf{x})} \mathcal{L}_{\rm adv}
\end{equation}

This objective is to minimize the adversarial loss to get the optimal generator and discriminator networks $(G^*, D^*)$. The {\tt arg} refers to the generator $G$ and discriminator $D$ to be optimized, $\mathbb{E}$ refers to the expectation of the adversarial loss over all $(\mathbf{z}, \mathbf{x})$, and {\tt min} and {\tt max} refers to updating the G parameters to minimize the expectation and updating D to maximize the expectation. In the unconditional case, the random vector $\mathbf{z}$ provides stochastic features, while in the conditional case, extra astrophysical parameters $\mathbf{c}$ serve as the condition for the generator. The real image samples are represented by $\mathbf{x}$, which is drawn from $p(\mathbf{x})$, imitated by the training set, while $p(\mathbf{z})$ is modeled as a multivariate diagonal Gaussian distribution.

For this work, we used StyleGAN2 \citep{Karras_2019} as our GAN architecture. The generator consists of two main parts: a mapping network $f$ and a synthesis network $g$. The mapping network takes the astrophysical parameters $\mathbf{c}$ and the random vector $\mathbf{z}$ as inputs and returns a style vector $\mathbf{w}$. The synthesis network uses the style vector $\mathbf{w}$ to adjust the weights in convolution kernels, while Gaussian random noise is added to the feature map after each convolution. The generator creates images with a low resolution, then progressively upsamples them until they reach the desired size. The discriminator, on the other hand, uses a ResNet \citep{He_2015} architecture. Our implementation of StyleGAN2 is adapted from this GitHub repository\footnote[1]{\url{https://github.com/dkn16/stylegan2-pytorch}, which is based on \url{https://github.com/rosinality/stylegan2-pytorch}.}.

We also apply regularization techniques to both the generator and discriminator components. Specifically, we use an $r_1$ loss function \citep{Lars_2018} to encourage sparsity in the discriminator's weight matrices, and a path-length loss function to encourage smoothness in the generator's output
\begin{equation}
    \mathcal{L_{\mathrm{path}}} = \left(\Big|\Big|\frac{\partial G(\mathbf{w})}{\partial \mathbf{w}}^{T}G(\mathbf{w})\Big|\Big|_2-a\right)^2,
\end{equation}
where $\mathbf{w}$ is the network weights, and $a$ is a constant to make the results more reliable and the model more consistently behaving.

The final training objective is a combination of these regularization terms and the original adversarial loss function. By including these regularization terms, we aim to prevent overfitting and improve the generalization ability of the model. Our final training objective is
\begin{equation}
    \begin{aligned}
       & (G^*,D^*) = \arg \min\limits_{G} \max \limits_{D} \mathbb{E}_{\mathbf{z}\sim p(\mathbf{z}),\mathbf{x}\sim p(\mathbf{x})} \mathcal{L}_{\rm adv}\\
    &+\arg \min\limits_{D}\mathbb{E}_{\mathbf{z}\sim p(\mathbf{z}),\mathbf{x}\sim p(\mathbf{x})}\mathcal{L}_{r_1}+\arg\min\limits_{G}\mathbb{E}_{\mathbf{z}\sim p(\mathbf{z}),\mathbf{x}\sim p(\mathbf{x})}\mathcal{L_{\mathrm{path}}}
    \end{aligned}
\end{equation}

\section{Statistics and Metric}
\label{sec:Sta_Met}

\subsection{Scattering transform}
\label{sec:ST}
Here we give a brief summary of the scattering transform based on \citep{ cheng2020new}. When applied to images, the scattering transform basically ``scatters'' the image information into different scales iteratively and outputs scattering coefficients that are informative summaries of the images. It is analogous to convolutional neural networks (CNNs): it contains similar convolution, non-linear function, and pooling operation, but it does not need the training process; It also shares some properties with $N-$point correlation functions: it is able to extract non-Gaussian information but does not amplify the fluctuations in the images. The basic scattering process is as follows. 
\begin{equation}
\begin{aligned}I_{0} & \equiv \text { input image } \\ I_{n}^{j, l} & \equiv\left|I_{n-1} \star \psi^{j, l}\right| \\S_{n} & \equiv\left\langle I_{n}\right\rangle\end{aligned}
\end{equation}
where $\psi^{j_{1}, l_{1}}$ is the wavelets parameterized by the scale $j$ and orientation $l$, $\star \psi^{j, l}$ is the covolution with wavelets, $|\cdot|$ is the modulus, and $\left\langle \cdot \right\rangle$ is the spatial average which outputs the scattering coefficients.
The commonly used zeroth-, first-, and second-order scattering coefficients can be formulated by
\begin{equation}
\begin{aligned} S_{0} & \equiv\left\langle I_{0}\right\rangle \\ S_{1}^{j_{1}, l_{1}} & \equiv\left\langle I_{1}^{j_{1}, l_{1}}\right\rangle=\left\langle\left|I_{0} \star \psi^{j_{1}, l_{1}}\right|\right\rangle \\ S_{2}^{j_{1}, l_{1}, j_{2}, l_{2}} & \equiv\left\langle I_{2}^{j_{1}, l_{1}, j_{2}, l_{2}}\right\rangle=\left\langle|| I_{0} \star \psi^{j_{1}, l_{1}}\left|\star \psi^{j_{2}, l_{2}}\right|\right\rangle\end{aligned}
\end{equation}
For our image with $64\times 64$ pixels, we use $J=5$ and $L=4$ so that $j_1\in\{0,\ldots, J-1\}$ and $l\in\{0,\ldots, L-1\}$. For the second-order coefficients, we only keep the informative coefficients with scale $j_2>j_1$ so that $j_2\in\{j_1+1,\ldots,J-1\}$ for $j_1<J-1$. We average over $l_1$ for the first-order scattering coefficients while keeping $l_2-l_1$ as an index and only averaging over $l_1$ for the second-order scattering coefficients. In this way, we reduce the number of coefficients but still preserve some morphological information. The final scattering coefficients have a dimension of 46. Because our scattering coefficients all have non-negative values\footnote[2]{Our simulation setting guarantees that there are no negative pixel values in the image. The well-trained generative models also produce non-negative values at most pixels. For other cases where the spatial average (the zeroth-order coefficient) is negative, one could take the logarithm of its amplitude and keep the sign, or just leave it untransformed.}, we follow \citet{cheng2020new} and take the logarithm with base 10 of all coefficients. All of the quantitative analyses in the following sections are based on these logarithmic coefficients. The package we used to calculate the scattering coefficients can be found in this link\footnote[3]{\url{https://github.com/SihaoCheng/scattering\_transform}}.

\subsection{Fréchet Scattering Distance}
\label{sec: fsd}

The Fréchet Inception Distance (FID, \citealp{NIPS2017_8a1d6947}) is used to evaluate the generative models. It takes the feature vectors from a pre-trained Inception-v3 \citep{7780677} model, a powerful CNN architecture, and calculates the Fréchet distance \citep{DOWSON1982450} between two groups of feature vectors, one group from the generated images by the generative model, and the other from the real or simulated images. On the one hand, the aforementioned inception model is trained with images like cats or dogs, so it is not suitable for our case. On the other hand, the scattering transform has a good analogy to conventional CNNs. So it is natural for us to construct a similar distance metric but with the Inception-v3 model replaced by scattering transform, dubbed Fréchet Scattering Distance (FSD)
\begin{equation}
d^{2}=\left|\mu_{X}-\mu_{Y}\right|^{2}+\operatorname{tr}\left(\Sigma_{X}+\Sigma_{Y}-2\left(\Sigma_{X} \Sigma_{Y}\right)^{1 / 2}\right),
\end{equation}
where $X$ and $Y$ are the scattering coefficients calculated from two samples of images, and $\mu$ and $\Sigma$ are the mean and covariance matrix, respectively. So these two terms in this equation represent the mean distance (summation of the mean difference square) and a distance defined in the covariance matrix space. The FSD can measure the similarity between two samples drawn from their own distributions, where an FSD closer to zero indicates more similarity. We calculate the FSD based on images generated from each generative model and simulation. Note that the scattering coefficients usually have much larger values for the zeroth- and first-order coefficients, which will make these components dominate the total FSD. The logarithm will bring different orders of coefficients to a similar level of magnitude. One can also consider rescaling different components to the same range, e.g.\ $[0,1]$, while we find that this rescaling leads to little difference in our conclusion. The FSD is used for both testing and evaluation during training.

\begin{figure*}
    \centering
    \includegraphics[width=\textwidth]{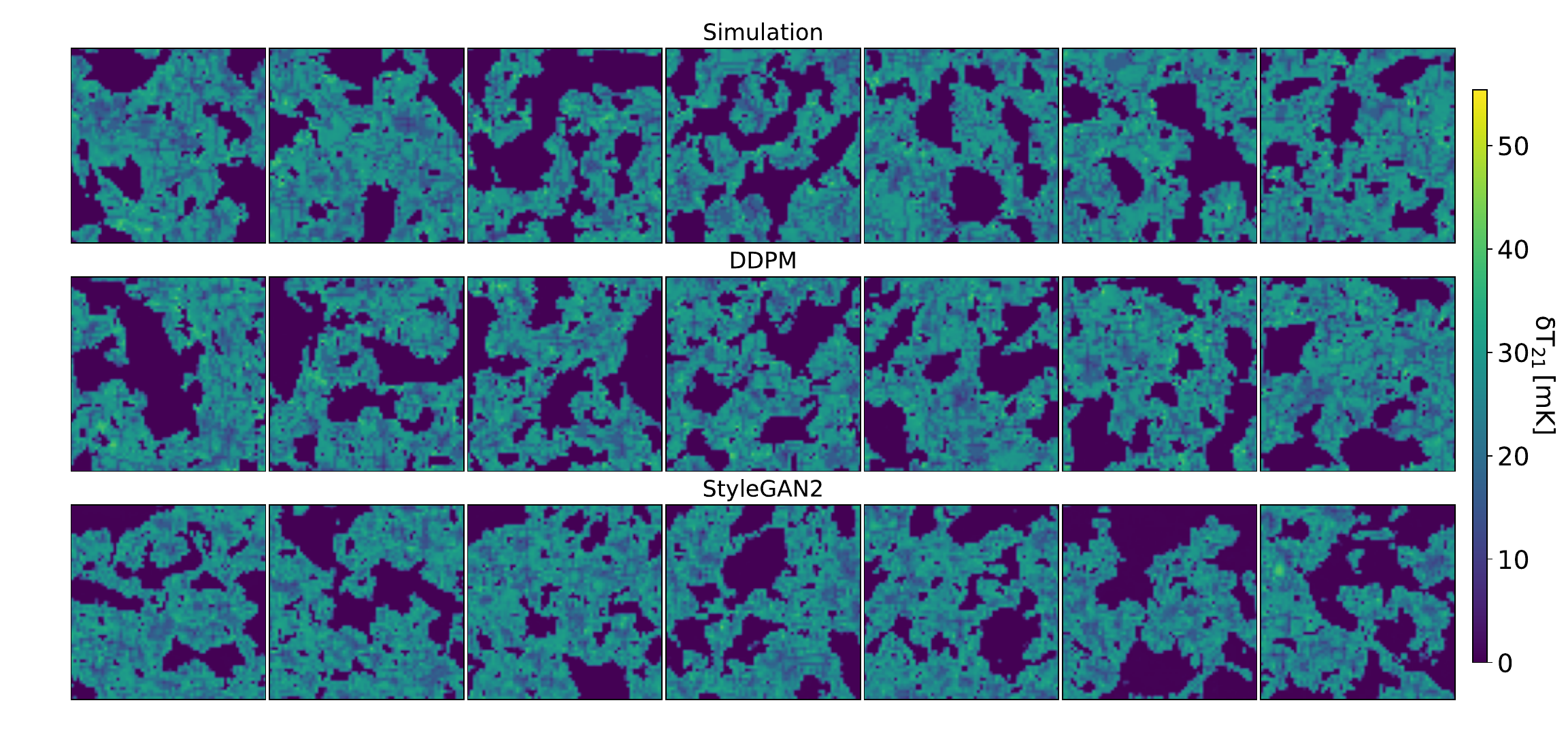}
    \caption{Conditional image generation from Simulation, DDPM, and StyleGAN2, at the testing set $(\mathrm {\tilde T } _ { \mathrm { vir } }, \tilde \zeta)=(0.200, 0.800)$.}
    \label{fig: cond_gen}
\end{figure*}

\section{Data preparation}
\label{sec:simu}
The 21~cm brightness temperature \citep{Furlanetto2006} relative to the CMB temperature $T_{\rm \gamma}$ at position $\textbf{x}$ is defined as
\begin{equation}
T_{21}(\textbf{x},z)=\tilde{T}_{21}(z)\,x_{\rm HI}(\textbf{x})\,\left[1+\delta(\textbf{x})\right]\,(1-\frac{T_{\rm \gamma}}{T_S})\,,\label{eqn:21cm}
\end{equation}
where $\tilde{T}_{21}(z) = 27\sqrt{[(1+z)/10](0.15/\Omega_{\rm m} h^2)}(\Omega_{\rm b} h^2/0.023)$ in units of mK. 
Here, $x_{\rm HI}({\bf x})$ is the neutral fraction, and $\delta({\bf x})$ is the matter overdensity, at position ${\bf x}$. Assuming that the baryon perturbation follows the cold dark matter on large scales, we can establish that $\delta_{\rho_{\rm H}} = \delta$. This paper also assumes the spin temperature $T_S$ greatly surpasses $T_{\rm \gamma}$, which is applicable soon after the onset of reionization. 

In this work, we use the semi-numerical code 21cmFAST \footnote[4]{\url{https://21cmfast.readthedocs.io/en/latest/}} v3.0.4 to generate our dataset. The 21cm brightness temperature box is simulated at the redshift $z \sim 11.93$, with a resolution of $64^3$ cells and box size of $(128 \ {\rm cMpc})^3$. The representative parameters we choose to vary are as follows:
\begin{itemize}
\item $\mathrm { T } _ { \mathrm { vir } }$, the minimum viral temperature of halos producing ionizing photons. In order to construct our dataset, we vary this parameter as $\log _ { 10 } \mathrm { T } _ { \mathrm { vir } }  \in [ 4, 6 ]$. For simplicity of training, it is further re-scaled to $ \mathrm {\tilde T } _ { \mathrm { vir } } \in [0,1]$.

\item $\zeta$, the ionizing efficiency of galaxies, which is a combination of several parameters related to ionizing photons. We vary this parameter as $\zeta \in [ 10,250 ]$, take the logarithm with base 10, and re-scale it to $\tilde \zeta \in [ 0,1 ]$.

\end{itemize}

The cosmological parameters we used are ( $\mathrm{\Omega _ { m }}$, $\mathrm{\Omega _ { b }}$, $\mathrm{n_s}$, $\mathrm{\sigma _ { 8 }}$, $\mathrm{h}$ ) = (0.310, 0.0490, 0.967, 0.810, 0.677) \citep{ade2016planck}.

\begin{figure*}
    \centering
    \includegraphics[width=\textwidth]{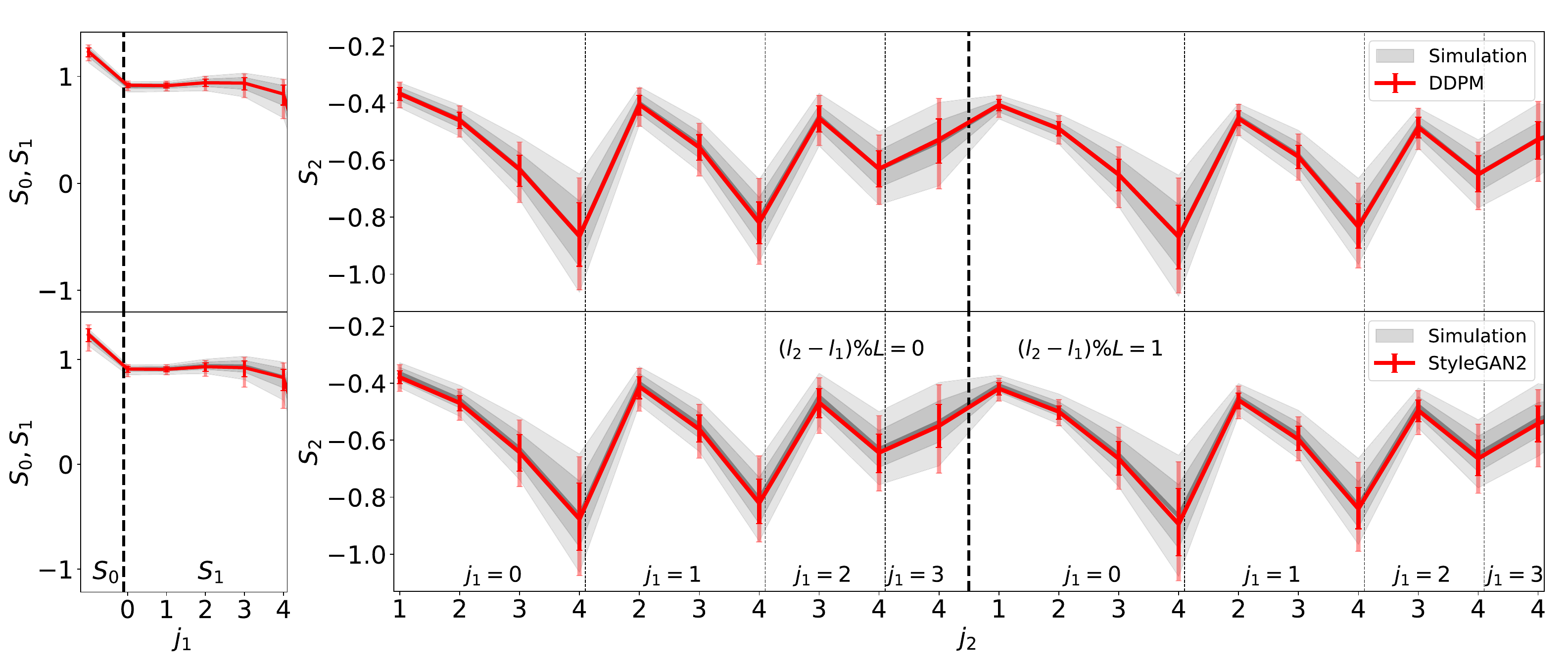}
    \caption{The reduced scattering coefficients over different scales $j$ and angular frequency $l$ for conditional image generation. These coefficients are calculated from images with a grid size of $(64,64)$ and have the wavelet parameters $J=5$ and $L=4$. The zeroth-order coefficient $S_0$ is the spatial average of the input image. The first-order coefficient $S_1(j_1)$ is the results averaged over orientation $l_1$. For each scale $j_1$, the second-order coefficient $S_2$ has the scale $j_2>j_1$ and is a function of the separation of two orientations $(l_2-l_1)\%L=\{0,1,2,3\}$. For simplicity, here we only show the coefficients with $(l_2-l_1)\%L=\{0,1\}$. For StyleGAN2, the $S_0$ shows a larger variance, and the $S_1$ and $S_2$ show a slightly larger bias in some scales than DDPM.}
    \label{fig: reduced_ST}
\end{figure*}

In order to construct a dataset covering most of the parameters space efficiently, we use the Latin Hypercube Sampling \citep{Mckay_2000} of parameters, each parameter with one realization of simulation box (from one random realization of the initial conditions). The reason we choose these one-to-one parameter-image pairs rather than one-to-multiple ones is that we want a dataset spanning most of the parameter space while keeping the computational need as low as possible. We choose a certain slice with the resolution of $64^2$ cells in each simulation box and a total of 25600 slices with the corresponding astrophysical parameters as our training dataset. For our testing dataset, considering the computational costs, we choose five reference points: 

\begin{enumerate}
    \item $(\mathrm {\tilde T } _ { \mathrm { vir } }, \tilde \zeta)=(0.200, 0.800)$ or $(\log_{10} \mathrm {T}_{\mathrm{vir}}, \zeta)=(4.400, 131.341)$, $(\mathrm {\tilde T } _ { \mathrm { vir } }, \tilde \zeta)=(0.800, 0.200)$ or $(\log_{10} \mathrm {T}_{\mathrm{vir}}, \zeta)=(5.600,19.037)$. These two sets of parameters represent two extremes where there are the most and least visual features in the corresponding 21~cm images.
    \item $(\mathrm {\tilde T } _ { \mathrm { vir } }, \tilde \zeta)=(0.351, 0.340)$ or $(\log_{10} \mathrm {T}_{\mathrm{vir}}, \zeta)=(4.699, 30.000)$, $(\mathrm {\tilde T } _ { \mathrm { vir } }, \tilde \zeta)=(0.739, 0.931)$ or $(\log_{10} \mathrm {T}_{\mathrm{vir}}, \zeta)=(5.477,200.000)$. These two sets of parameters correspond to the ``Faint Galaxy Model'' and ``Bright Galaxy Model'' in \citet{2017MNRAS.472.2651G}. The corresponding images have modest features.
    \item $(\mathrm {\tilde T } _ { \mathrm { vir } }, \tilde \zeta)=(0.400, 0.800)$ or $(\log_{10} \mathrm {T}_{\mathrm{vir}}, \zeta)=(4.800,131.341)$, another parameter set which also has obvious features in the images.
\end{enumerate}

These five parameter sets are representative in the parameter space. At each point, we generate 800 realizations with varying initial conditions (4000 testing images in total). We have verified that this number of realizations is enough to calculate the sample variance. For further analysis, we also generate samples around two of these testing sets, which will be discussed in Section~\ref{sec:fisher}. 

\section{Results}
\label{sec: results}

For astrophysical applications, image generation should be made conditional on some information in need, e.g.\  initial conditions, cosmological or astrophysical parameters, so that we can perform the Bayesian inference of these parameters from observations. Thus, we focus on the comparison of conditional image generations between the two generative models.

 We train the conditional generative models on several datasets with increasing sample sizes. After training, we generate 800 images conditional on the five testing sets. Then we compare the distribution of the scattering coefficients calculated from the generated images with that from the simulations at each set. We make the comparison both qualitatively and quantitatively by using FSD and Fisher forecasts.

\subsection{FSD performance}
\label{sec: results_fsd}

\begin{table}
	\centering
	\caption{Average FSD over five testing sets for varied training sample size $(1600-25600)$, where DDPM (0) means zero guidance scale, i.e. $s=0$ in the classifier-free guidance.}
	\begin{tabular}{cccccc} 
		\hline
		\hline
            Models    & &  & FSD & &  \\ \cline{2-6}
		        & 1600 & 3200 & 6400 & 12800 & 25600 \\
		\hline
		StyleGAN2    & 0.0371& 0.0263& 0.0213& 0.0180& 0.0160\\
		DDPM (0)   & $\boldsymbol{0.0117}$ & $\boldsymbol{0.0057}$ & $\boldsymbol{0.0037}$ &$\boldsymbol{0.0023}$ &$\boldsymbol{0.0024}$\\
		\hline
		\hline
	\end{tabular}

	\label{tab:cond}
\end{table}

\begin{figure*}
    \centering
    \includegraphics[width=\textwidth]{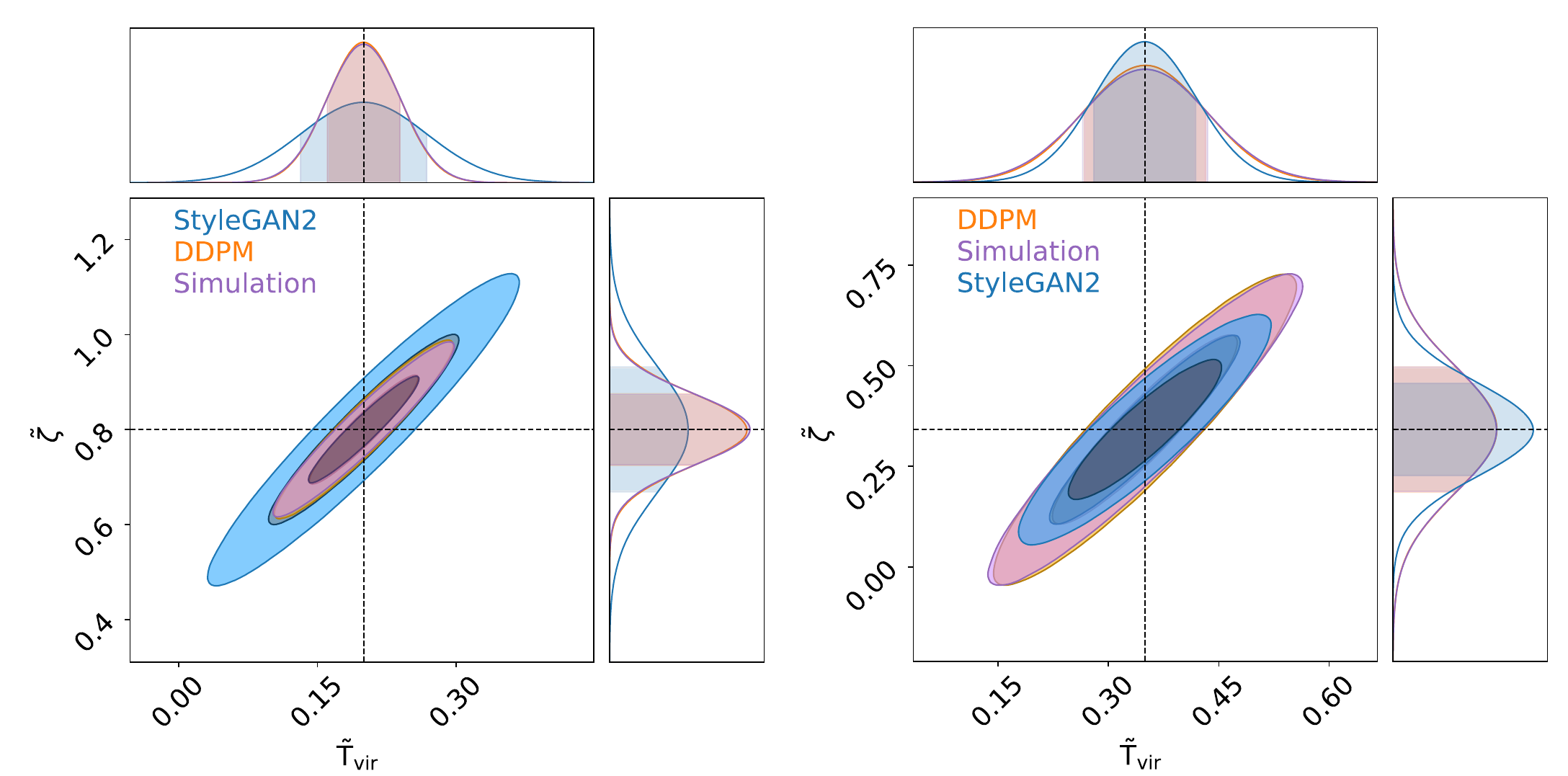}
    \caption{Fisher forecasts of two sets of astrophysical parameters which are re-scaled to [0,1]. The true values, $(\mathrm {\tilde T } _ { \mathrm { vir } }, \tilde \zeta)=(0.200,0.800)$ (left) and $(\mathrm {\tilde T } _ { \mathrm { vir } }, \tilde \zeta)=(0.350,0.341)$ (right), are indicated by the black dash lines. The DDPM accurately recovers the distribution of the parameters, whereas StyleGAN2 significantly biases the constraints.}
    \label{fig: fisher}
\end{figure*}

We visualize the image generations conditional on one testing set $(\mathrm {\tilde T } _ { \mathrm { vir } }, \tilde \zeta)=(0.200,0.800)$ in Fig.~\ref{fig: cond_gen}, based on 25600 training samples\footnote[5]{We use the ChainConsumer, \url{https://github.com/Samreay/ChainConsumer}}. The images from the generative models have some small negative values, while here we make a cutoff at zero for better visualization. The reduced scattering coefficients calculated from images at the same testing set are plotted in Fig.~\ref{fig: reduced_ST}, where the zeroth-order scattering coefficients from StyleGAN2 show a larger variance and the first- and second-order coefficients show a slightly larger bias in some scales than DDPM. The FSD of StyleGAN2 and DDPM are $0.0120$ and $0.0022$, respectively. The reference FSD is $0.0017$, which is calculated based on two groups of simulations at the same testing set. These results imply that StyleGAN2 learned biased astrophysical parameter dependence of images. Although our training dataset contains one-to-one parameter-to-image pairs, the DDPM still learns the variance on a specific parameter point accurately. This is possible because, in the training sample, different images are simulated from the initial conditions with the same distribution. 

In Table~\ref{tab:cond}, we show the FSD from StyleGAN2 and DDPM averaging over our five testing sets for generality, where the reference average FSD is $0.0018$. We find that the FSD improves slowly when increasing the training sample size. The DDPM outperforms StyleGAN2 even without any guidance ($s=0$), and the FSD from DDPM is always smaller than that from StyleGAN2 in terms of the different sizes of the training datasets.

\subsection{Fisher forcasts}
\label{sec:fisher}
We further explore the conditional generative performance by Fisher forecasts on the astrophysical parameters. The Fisher matrix \citep{10.2307/2342435, 2018PhRvD..97h3004C} can be written by
\begin{equation}
\mathbf{F}_{\alpha \beta}(\boldsymbol{\boldsymbol{c}})=-\left.\left\langle\frac{\partial^{2} \ln \mathcal{L}(\mathbf{d} \mid \boldsymbol{\boldsymbol{c}})}{\partial \boldsymbol{c}_{\alpha} \partial \boldsymbol{c}_{\beta}}\right\rangle\right|_{\boldsymbol{c}=\boldsymbol{c}^{\mathrm{fid}}}=\boldsymbol{\mu},_{\alpha}^{T} \mathbf{C}^{-1} \boldsymbol{\mu},_{\beta},
\end{equation}
where $\mathcal{L}(\mathbf{d} \mid \boldsymbol{\boldsymbol{c}})$ is the likelihood function, $\mathbf{C}$ is the covariance matrix which only contains sample variance in our case, $\boldsymbol{\mu},_{\alpha}$ and $\boldsymbol{\mu},_{\beta}$ are the derivatives with respect to the parameters. Here we approximate this derivative by calculating the difference quotient averaged over different realizations.
The Fisher matrix gives the minimum variance of an estimator of a parameter $\boldsymbol{c}$ (Cramér-Rao
bound, \citealp{radhakrishna1945information})
\begin{equation}
\left\langle\left(\boldsymbol{c}_{\alpha}-\left\langle\boldsymbol{c}_{\alpha}\right\rangle\right)\left(\boldsymbol{c}_{\beta}-\left\langle\boldsymbol{c}_{\beta}\right\rangle\right)\right\rangle \geq\left(\mathbf{F}^{-1}\right)_{\alpha \beta}
\end{equation}

\begin{figure}
    \centering
    \includegraphics[width=\linewidth]{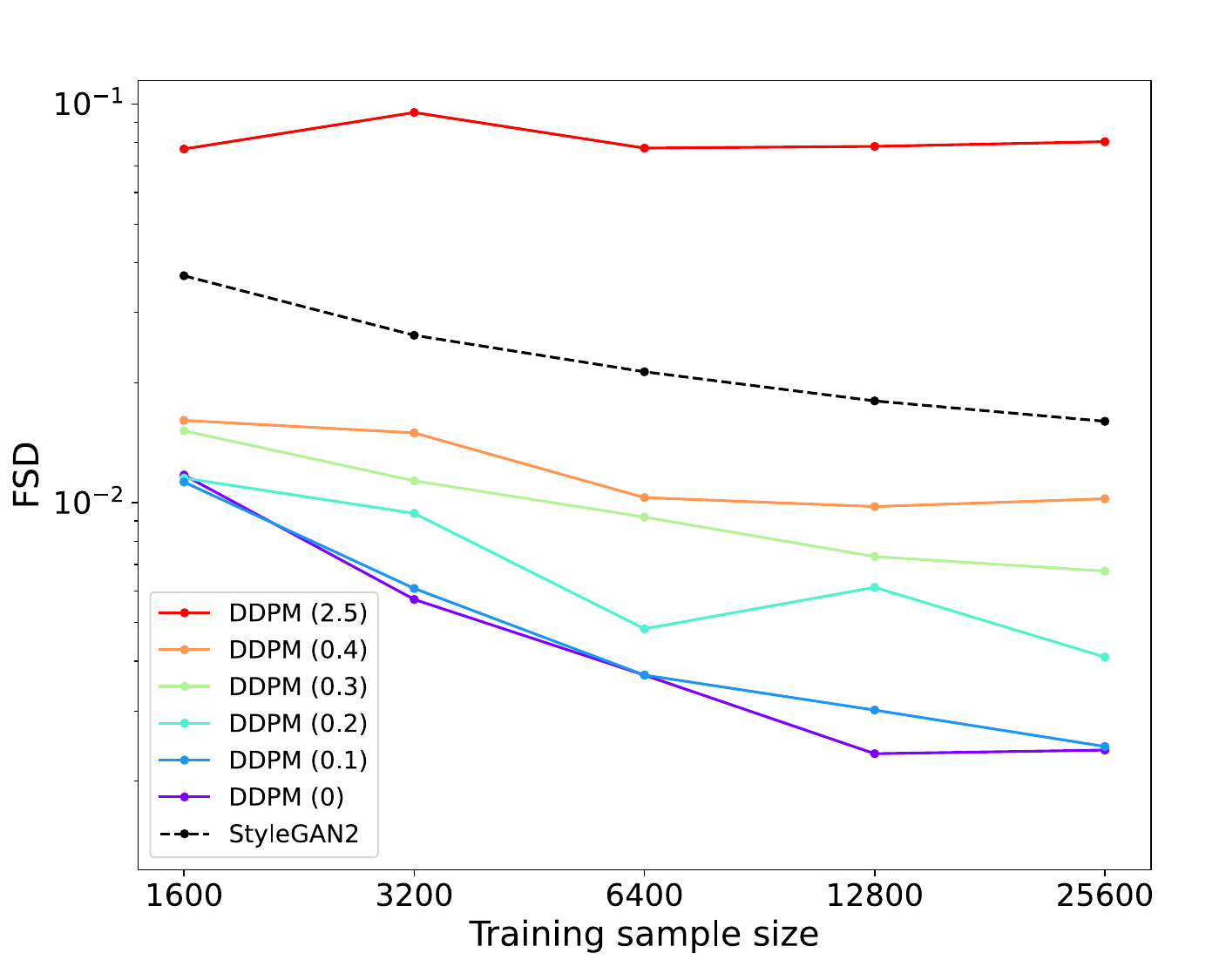}
    \caption{FSD as a function of training sample sizes with different guidance scales for DDPM, e.g. DDPM (0) means guidance scale $s=0$. A small positive guidance scale leads to a small FSD score only when the sample size is small (i.e.\ 1600).}
    \label{fig: guidance}
\end{figure}

\begin{figure*}
    \centering
    \includegraphics[width=\textwidth]{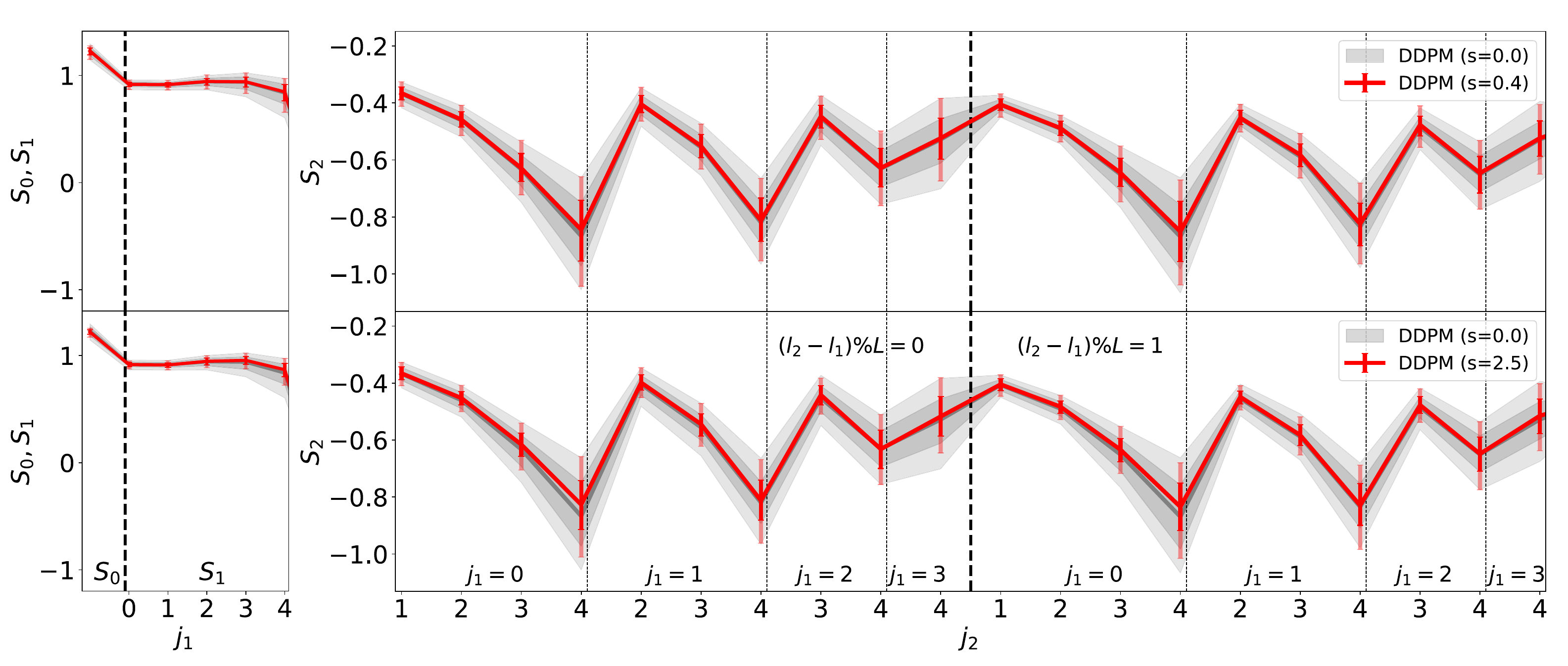}
    \caption{Same as Fig.~\ref{fig: reduced_ST} but comparing the DDPM sampling at three particular guidance scales $s$. The increasing guidance scales lead to mode collapse, where the reduced scattering coefficients have more biased means and less variance.}
    \label{fig: S theta scale}
\end{figure*}

This parameter covariance together with the ``best estimate'' of the parameters can give a Gaussian approximation of the true posteriors. In this work, we only focus on the estimated parameter uncertainty, so the best estimate is chosen to be the fiducial parameter at each testing set. The Fisher forecasting technique is also applied to constrain astrophysical parameters with statistical summaries for 21 cm mock observations \citep{2017MNRAS.468.1542S}. In this work, we omit the observational effects and adopt Fisher forecasting as a complement aspect to verify the performance of generative models, under the mild Gaussian approximation of scattering coefficients distribution (see Section~\ref{sec: assump}).

For the two generative models, we use the maximum training sample size of 25600 and calculate the Fisher matrix at the two testing sets $(\mathrm {\tilde T } _ { \mathrm { vir } }, \tilde \zeta)=(0.200.0.800)$ and $(\mathrm {\tilde T } _ { \mathrm { vir } }, \tilde \zeta)=(0.350,0.341)$, respectively. The second one produces images with less evolution (from a Gaussian density field) or fewer visual features. For each of these two points, we use 800 realizations for covariance calculation and 400 realizations for derivative calculation, where the realizations are controlled by the random number generator {\tt torch.manual\_seed}. The parameter step size to calculate the finite difference is $\pm 0.0025$ for both points, we have checked that both Fisher matrix has converged with this parameter step size. 

The results are shown in Fig.~\ref{fig: fisher}, where the DDPM gives comparable parameter constraints to simulations at both testing sets, while StyleGAN2 performs worse on either point. We use the reciprocal of the area of the $1\sigma$ Fisher forecast ellipse as the figure of merit (FoM, \citealp{2009arXiv0906.4123C}). The error of FoM for DDPM is $9\%$ and $4\%$ at the two points, while for StyleGAN2, the error becomes $69\%$ and $33\%$, indicating that StyleGAN2 does not learn the parameter dependence well or does not recover the image distribution conditional on the specific parameters, which are signs of the well-known partial mode collapse problem for GANs. We give a more detailed analysis in Section~\ref{sec: fisher_GAN}, where we show the GAN either learned fewer changes around the first testing set or learned less variance around the second testing set, verifying the model collapse hypothesis. 

\section{Discussions}
\label{sec: discuss}

\subsection{Effect of guidance scale for DDPM}
\label{sec: guidance}
The classifier-free guidance is reported to give a smaller FID \citep{ho2021classifier} when applied to the class-conditional generation of natural images if choosing a proper non-zero guidance scale. However, our similar metric—FSD applied to the parameter-conditional generation of astrophysical images shows an optimum at zero guidance scale for most experiments. In Fig.~\ref{fig: guidance}, we show the FSD as a function of training sample size with different guidance scales. We find that a large guidance scale results in a larger (worse) FSD. This trend is similar to the literatures using FID on natural images where larger guidance scale make the generated images more typical while sacrificing diversity. We also find that a small guidance scale leads to a better FSD only when the sample size is small (1600 in our case). Using classifier-free guidance has little gains when the training sample size is large. In Fig.~\ref{fig: S theta scale}, We show the typical dependence of the guidance scale on the reduced scattering coefficients. We find that with increasing guidance scale, the means of some components in scattering coefficients get more biased but the variance becomes smaller. Quantitatively, for DDPM trained on 25600 samples, we select 3 guidance scales: $\{0, 0.4, 2.5\}$. The mean distance in FSD, $\left|\mu_{X}-\mu_{Y}\right|^{2}$, is $\{0.8, 2, 9\}\times10^{-3}$. We use the trace of the covariance matrix $\mathrm{Tr}\left [ \mathrm{Cov}\ \mathbf{S}\right]$ of these scattering coefficients as the metric to approximate the absolute variance. The $\mathrm{Tr}\left [ \mathrm{Cov}\ \mathbf{S}\right]$ for these two scales is $\{6.5, 4.8, 3.0\}$. These results confirm that classifier-free guidance leads to mode collapse, where the generated images are biased and overly confident. Similar behavior of the classifier-free guidance is also found in \citet{2023arXiv230110677P}, where the authors used the diffusion model to fit the actions conditional on the observations in order to imitate human behavior. They found that the classifier-free guidance leads to less usual actions that are paired with some observation. 

\begin{figure*}
    \centering
    \includegraphics[width=\textwidth]{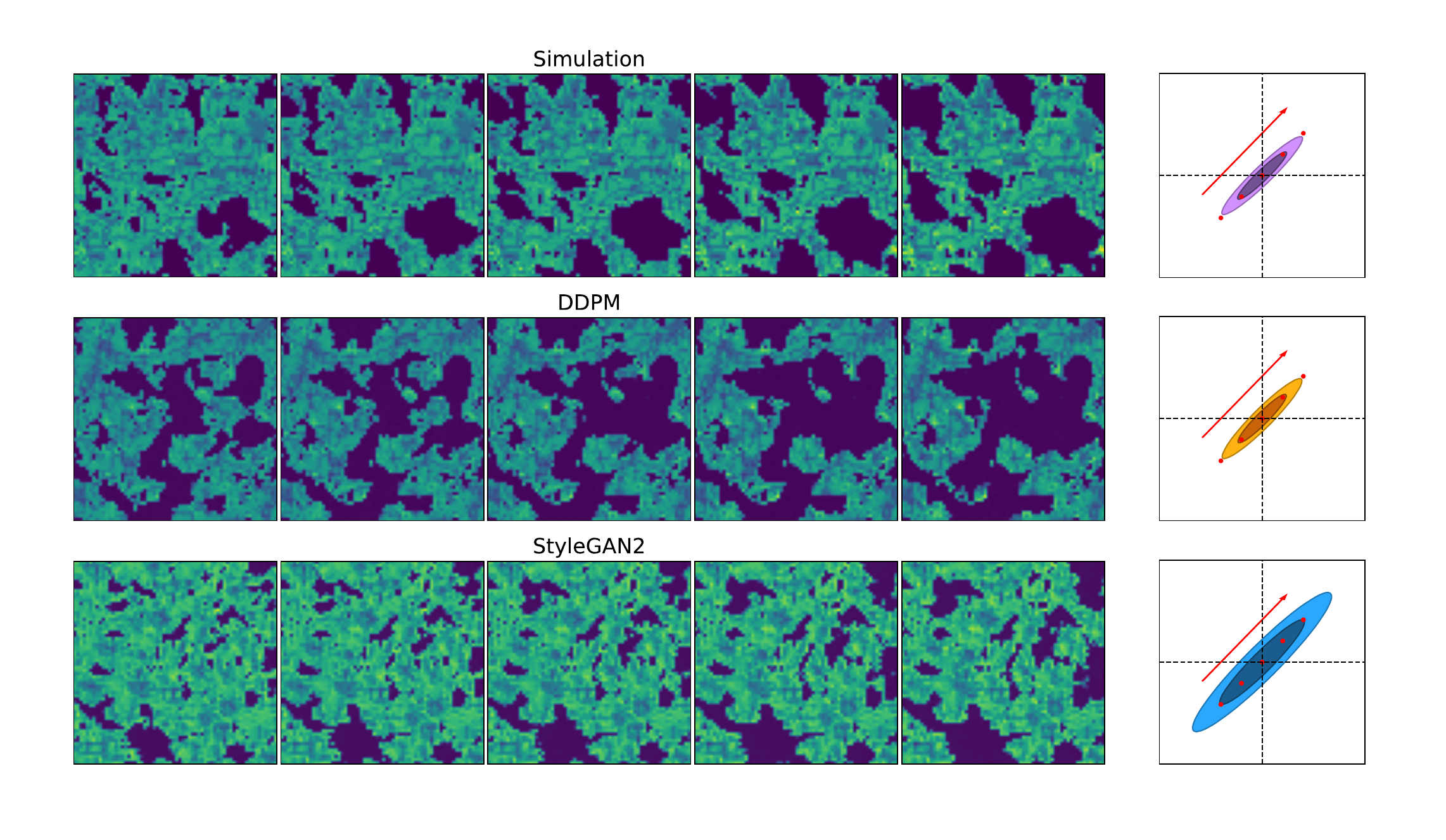}
    \caption{Conditional image generation by varying the parameters around the testing set $(\mathrm {\tilde T } _ { \mathrm { vir } }, \tilde \zeta)=(0.200,0.800)$. The images on the left panel are generated with equally spaced five parameter points roughly along the major axis of the credible parameter region in the Fisher forecasts (as indicated by the arrow in the right panel). The images from StyleGAN2 have the least changes, indicated by the global brightness temperature $\left \langle \mathbf{T}_{21} \right \rangle$, compared with simulations and DDPM.}
    \label{fig: S_param1_degeneracy_vary}
\end{figure*}

\subsection{Fisher forecasts by StyleGAN2}
\label{sec: fisher_GAN}
For the results of Fisher forecasts by StyleGAN2, we hypothesize that the overestimate and underestimate of the uncertainty at the two testing sets are different forms of the mode collapse. For the first testing set $(\mathrm {\tilde T } _ { \mathrm { vir } }, \tilde \zeta)=(0.200,0.800)$, The overestimate of the uncertainty, especially along the major axis, may indicate that there are fewer image changes along this direction in the 2d parameter space. In order to check that, in Fig.~\ref{fig: S_param1_degeneracy_vary}, we choose equally-spaced five points roughly along the major axis of the Fisher forecasts and generate one image at each point for simulation, DDPM, and StyleGAN2. We use a unique colormap for each group of images to visualize the changes more clearly. In order to construct a global single-value metric to indicate the speed of changes, we calculate the global (average) brightness temperature $\left \langle \mathbf{T}_{21} \right \rangle$ over all pixels in an image, fit a linear line with the least-squares regression over the five $\left \langle \mathbf{T}_{21}\right \rangle$ values and extract the slope of the regression line. Let $A \equiv d\left \langle\mathbf{T}_{21}\right \rangle/d \boldsymbol{c}|_{\text{Simulation}}$, $d\left \langle\mathbf{T}_{21}\right \rangle/d \boldsymbol{c}|_{\{\text{StyleGAN2}, \text{DDPM}\}} \sim \{0.82, 1.07\}A$. The images from StyleGAN2 do have the least changes compared with the other two, implying that StyleGAN2 learned similar modes at least around this testing set. We also note in Fig.~\ref{fig: S_param1_degeneracy_vary} that although the DDPM learned well the ``statistical evolution'' of patterns with varied different astrophysical parameters, in some specific regions of images generated by both models there is no or little visual change with the parameters, which is not physical. 

\begin{figure*}
    \centering
    \includegraphics[width=\textwidth]{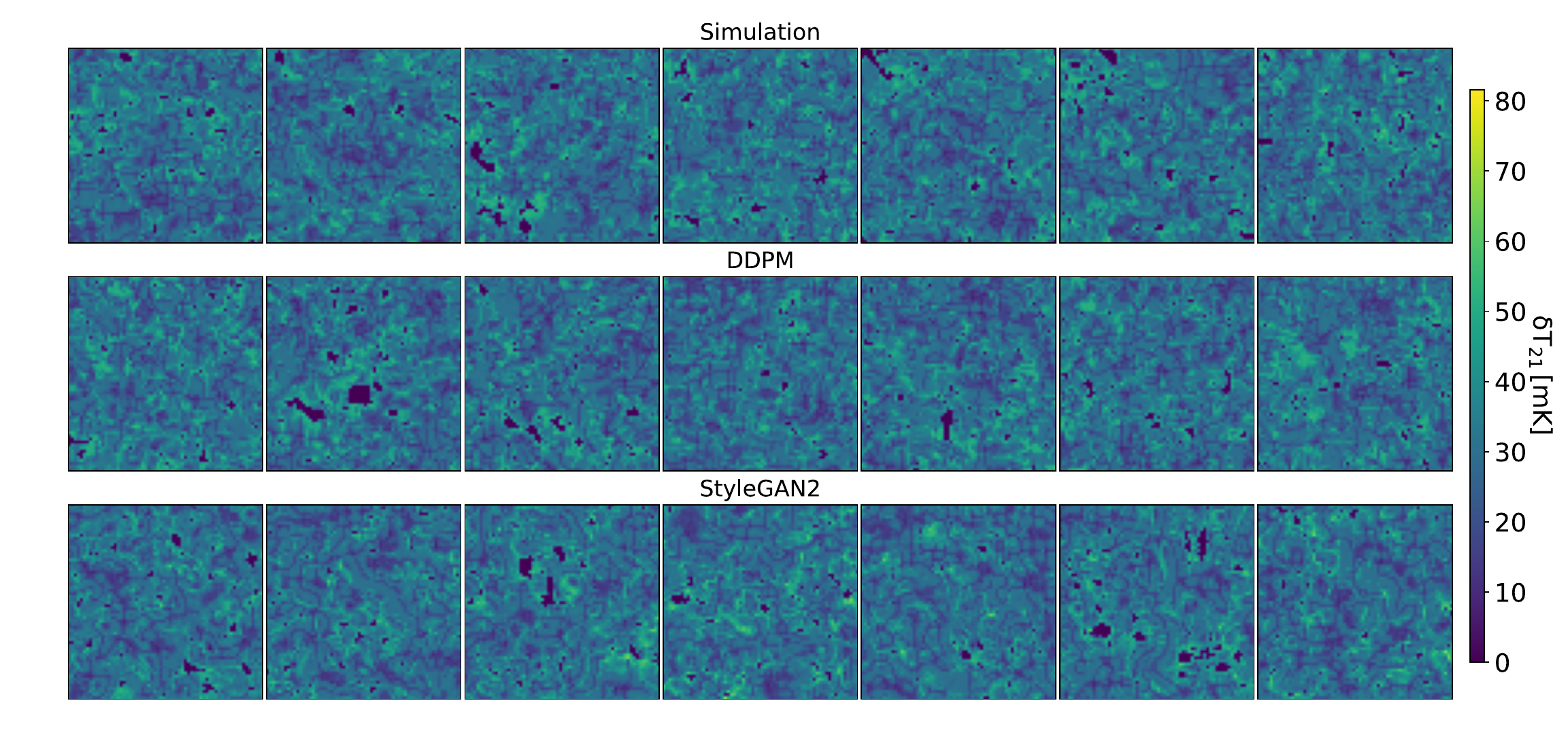}
    \caption{Same as Fig.~\ref{fig: cond_gen} but at the testing set $(\mathrm {\tilde T } _ { \mathrm { vir } }, \tilde \zeta)=(0.350,0.341)$.}
    \label{fig: image other}
\end{figure*}

\begin{figure*}
    \centering
    \includegraphics[width=\textwidth]{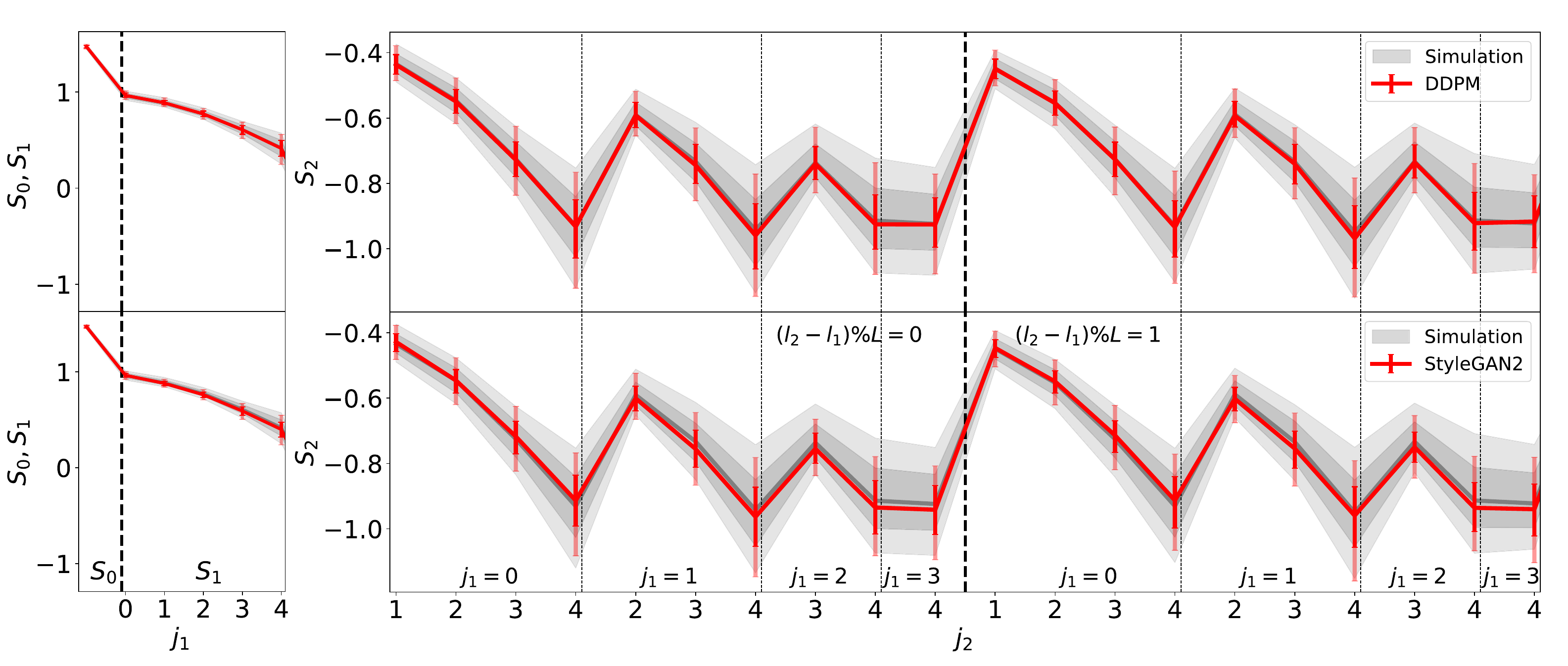}
    \caption{Same as Fig.~\ref{fig: reduced_ST} but at the testing set $(\mathrm {\tilde T } _ { \mathrm { vir } }, \tilde \zeta)=(0.350,0.341)$. The StyleGAN2 learns less variation and more biased means of some components than DDPM and simulations.}
    \label{fig: S other}
\end{figure*}

The underestimate of the uncertainty at the second testing set $(\mathrm {\tilde T } _ { \mathrm { vir } }, \tilde \zeta)=(0.350,0.341)$ may indicate that StyleGAN2 learned little variance there. In Fig.~\ref{fig: image other} and Fig.~\ref{fig: S other}, we show the generated images and the reduced scattering coefficients at the second testing set, respectively. Let $B \equiv \mathrm{Tr}\left [ \mathrm{Cov}\ \mathbf{S}\right]|_{\text{Simulation}}$, we find $\mathrm{Tr}\left [ \mathrm{Cov}\ \mathbf{S}\right]|_{\{\text{StyleGAN2}, \text{DDPM}\}}\sim \{0.70, 0.89\}B$. These results imply that StyleGAN2 does learn less variation for this specific astrophysical parameter. In addition to the smaller variance, StyleGAN2 also learned biased means of some components, indicating that when using StyleGAN2 as an emulator to perform parameter inference the results could be biased.

\subsection{Robust quantification of the generative model through scattering transform}
\subsubsection{Quantification of an astrophysical image}
\label{sec: stat}
This work relies on the summary statistics for two tasks: the evaluation of the image quality and the Fisher forecasts of the astrophysical parameters, where the latter is related to the broader topic of parameter inference. The key point in both tasks is the quantification of the features in an image. For images from the biological world where the non-Gaussian information is dominant, the features from CNNs are powerful to characterize their information. For a Gaussian field, the power spectrum is sufficient for the quantification. The astrophysical images lie between the random Gaussian field and the images in the biological world. In this context, the scattering transform has been an option to quantify the images. It has been shown that the scattering transform can reach comparable performance to the CNNs but with a more stable process \citep{cheng2020new}. Some moment-based statistics such as 3-point correlation function \citep{2019MNRAS.487.3050H,2020MNRAS.498.4518J} and its Fourier-space counterpart bispectrum \citep{2017MNRAS.468.1542S, 2022MNRAS.510.3838W} are also widely used to quantify the non-Gaussian information in the astrophysical fields. Compared with these statistics, the n-order scattering transform includes information up to $2^n$-point correlation function \citep{mallat2012group}. In addition, the scattering transform uses the non-expansive operator (modulus) and has low variance in the resulting coefficients, insensitive to the outliers in the images \cite{mallat2012group,6522407}. Therefore, the scattering transform up to the order of two is theoretically more stringent than (or at least equivalent to) the bispectrum. Empirically, the scattering transform also shows superiority over moment-based statistics such as power spectrum and bispectrum \citep{2021arXiv210309247C}, and 3-point correlation function \citep{2021ApJ...910..122S} for several astrophysical fields. Although what we use here is not totally full characterization, the use of more robust statistics has revealed limitations in using GAN for SBI. While GAN-generated cosmology can pass basic tests, such as the power spectrum \citep{10.1093/mnras/staa523}, as revealed by the scattering transform and our new evaluation metric FSD, DDPM recovers better sample distribution than StyleGAN2.

\subsubsection{Gaussian assumption in Fréchet distance}
\label{sec: assump}
Although the FID is among the most popular evaluation metrics for generative models, the Fréchet distance itself in particular holds for two Gaussian distributions. Actually, the features from the Inception models are shown not to be multivariate Gaussian \citep{BORJI2022103329}. In this work, we build upon the FID and develop our FSD which is effective in evaluating our two generative models. The scattering coefficients from the weak lensing convergence maps are shown to be well-Gaussianized \citep{2021arXiv210309247C}. The 21 cm images have different kinds of features and are potentially more complex than the lensing mass maps. So we have a detailed check of the Gaussianity of the scattering coefficients we obtained. For the 800 samples at the testing set $(\mathrm {\tilde T } _ { \mathrm { vir } }, \tilde \zeta)=(0.200,0.800)$, we apply the skewness test and find that for a significance level of 0.01, 25 out of the total 46 components of the scattering coefficients match a Gaussian distribution, which does not violate the Gaussian assumption much. This approximate Gaussianity contributes to the effectiveness of using scattering transform in this work. Considering the non-Gaussian part of the features, metrics like the kernel inception distance (KID,  \citealp{bińkowski2018demystifying}) could also be implemented in evaluating astrophysical images.  

\subsection{Early stopping}
\label{sec: stop}
A proper way to evaluate the generative models could be helpful for StyleGAN2 to notice the mode collapse or unstable training and to early stop the training. For DDPM, as specified in Appendix~\ref{sec: net}, we monitor the FSD based on our first testing $(\mathrm {\tilde T } _ { \mathrm { vir } }, \tilde \zeta)=(0.200,0.800)$ during training and choose the models that have the smallest FSD at that testing set. The goodness at a specific testing set is necessary but not sufficient for a good generative model. One can instead monitor the FSD at multiple testing sets. When monitoring the FSD at the same single testing set, as we do for DDPM, the average FSD from StyleGAN2 gets increasingly worse at larger sizes of training data. Upon investigation, we find that this average FSD is dominated by large FSD values at the testing sets where the corresponding images have modest or the least features, implying partial mode collapse to the parameter space with clearer image features. Finally, we choose to monitor all five testing sets for StyleGAN2, choose the model that performs well at all five testing sets, and report the results in Table~\ref{tab:cond}. In practice, we pre-set a maximum number of iterations, run all of the iterations, and pick the checkpoint of the model with the least FSD. We find that this maximum iteration is a conservative choice and the training can be early stopped when the FSD tends to be stable (or starts to get worse at some iteration for StyleGAN2).

\subsection{Limitations}
\subsubsection{Limitation of the dataset}
\label{sec: dataset}
Our training data-21 cm slices are pinned on a high redshift where a significant part of the astrophysical parameter space has 21 cm images with little evolution so that it is close to the Gaussian density field, in which case it is harder for the generative models to learn meaningful parameter dependence and calculate the accurate Fisher forecasts with the summary statistics from generated images. To make the most use of the diffusion model as an emulator, a better choice can be constructing a dataset that contains the evolution over a large redshift range, which makes the signal more distinguishable. Another consideration in constructing the training dataset is to explicitly include multiple realizations centered at each parameter, which should make it easier for the generative model to fit the joint parameter-image distribution, at the cost of more expensive data computation.

As a proof of concept, we operate with a limited dataset in terms of both sample size and image dimensions. Nevertheless, the extension to larger images can be readily achieved by encoding the extensive images into smaller counterparts utilizing an auto-encoder \citep{rombach2021highresolution} or by employing some useful tricks \citep{2023arXiv230111093H} to facilitate end-to-end high-resolution image training and generation.

\subsubsection{Sampling costs}
\label{sec: cost}
The sampling speed (on a single Nvidia Tesla V100 GPU) of DDPM is about $160$ times slower than StyleGAN2 and about $5$ times faster than the simulation (optimized on CPU). The reason for the slower sampling speed is that in our simplest implementation, the sampling will go through a Markov Chain for all time steps ($T=1000$ in our case), starting from the random noise to the final clean image. This process will call the trained U-Net 1000 times and is computationally expensive. However, one can also construct a non-Marcov chain for the diffusion process and then the reverse sampling of the diffusion process can be much faster, which is the basic idea of the denoising diffusion implicit model (DDIM, \citealp{songdenoising}). The DDIM has the same training procedure as DDPM but is $10-50$ times faster in sampling. There are also some other efforts to overcome the slow sampling process like applying consistency models \citep{2023arXiv230301469S} which can distill the pre-trained diffusion model like ours and achieve fast one- or few-step sampling. 

\subsubsection{Benchmark of generative models}
In this work, we put our focus on comparing two typical members, the DDPM and StyleGAN2 from the group of diffusion models and GANs, respectively. Specifically, we choose StyleGAN2 for its usage frequency in the astronomy community. It would also be interesting to compare DDPM with other GAN models which show superiority in terms of the FID over StyleGAN2 \citep{Karras2020ada,Sauer2021NEURIPS, Karras2021,10.1145/3528233.3530738}, while we note that for our maximum budget of training data, DDPM has the FSD about 7 times smaller than StyleGAN2, which is a significant improvement compared with the FID improvement of these GAN models over StyleGAN2 when applied to natural images, despite the fact that these GAN models operate on a totally different image space and effectiveness of this comparison might be limited. On the other hand, the DDPM architecture we use here is also not optimal in the diffusion model community. The current state-of-the-art generative models on various datasets are dominated by cutting-edge diffusion models \citep{2022arXiv220611474L, 2022arXiv221117091K, 2023arXiv230314389G}

\section{Conclusion}
\label{sec:conclusion}
Within our investigation, we we aim to assess the ability of the diffusion model to conditionally generate astrophysical images, and undertake a comparison between DDPM, a typical diffusion model, and StyleGAN2, focusing on their performance in generating 21 cm astrophysical images based on specific astrophysical parameters. We propose the utilization of the scattering transform, a more robust summary statistic which has been developed for cosmological studies to evaluate the precision and reliability of SBI. To enable a unified assessment of the image generations, we formulate the FSD by substituting the representations from the Inception-v3 model in FID with the scattering coefficients.

Our evaluation methodologies unveil that DDPM achieves superior recovery of image distributions compared to StyleGAN2. Notably, the distribution attained by DDPM exhibits remarkable proximity to the underlying true distribution derived from simulations, despite the networks being exposed solely to one image realization per parameter. Moreover, our results show that the use of classifier-free guidance benefits image generation only when the training sample is limited and the guidance scale is small. Otherwise, the classifier-free guidance results in the emergence of mode collapse and should be avoided when enough training data is available. A possible explanation suggested by our quantitative results is that when given more training data, the DDPM is more well-trained and can sample images with high quality without the need of introducing classifier-free guidance which was initially proposed to reach higher-quality image sampling while sacrificing the image diversity.

To delve deeper into the potential of generative models as emulators for parameter inference, we undertake Fisher forecasts based on the trained generative models. Our findings demonstrate that DDPM can provide parameter constraints comparable to those obtained from simulations across different testing sets, whereas StyleGAN2 manifests the issue of mode collapse, evident in either the parameter dependence of generated images or the conditional image distribution with respect to testing parameters.

Serving the purpose of verifying the performance of generative models compared with simulations, our investigations are conducted under circumstances where observational effects such as the thermal noise and foreground are not included throughout the analysis. However, the observational effects can be directly applied on top of the generated images for further downstream tasks. With such additional consideration, the diffusion model exemplifies the potential to serve as an emulator for astrophysical image generation, seamlessly integrated into a classical Monte Carlo Markov Chain \citep{Greig2015} for astrophysical parameter inference. It also showcases the prospect of performing SBI by learning (score of) the posterior, as well as the score of the likelihood, employing DDPM and its score-based variants \citep{2022arXiv221004872S} to infer astrophysical parameters.

\section*{Acknowledgements}

This work was supported by the National SKA Program of China (grant No.~2020SKA0110401), NSFC (grant No.~11821303), and National Key R\&D Program of China (grant No.~2018YFA0404502). YST acknowledges financial support from the Australian Research Council through DECRA Fellowship DE220101520. We thank Paulo Montero-Camacho for the useful discussions. We acknowledge the Tsinghua Astrophysics High-Performance Computing platform at Tsinghua University and the integrated high-performance research computing environment HCI in Australia, for providing computational and data storage resources that have contributed to the research results reported within this paper.

\section*{Data Availability}
The code for our implementation of DDPM is available at  \url{https://github.com/Xiaosheng-Zhao/ST4Diffusion}; the code for our implementation of StyleGAN2 is available at \url{https://github.com/dkn16/stylegan2-pytorch}. The derived data used in this research will be shared upon reasonable request to the corresponding author.



\bibliographystyle{mnras}
\bibliography{Ref} 



\appendix

\section{Network and hyperparameter details}
\label{sec: net}

\subsection{U-Net}
\label{sec: unet}

\begin{figure*}
\centering
 \includegraphics[width=.6\linewidth]{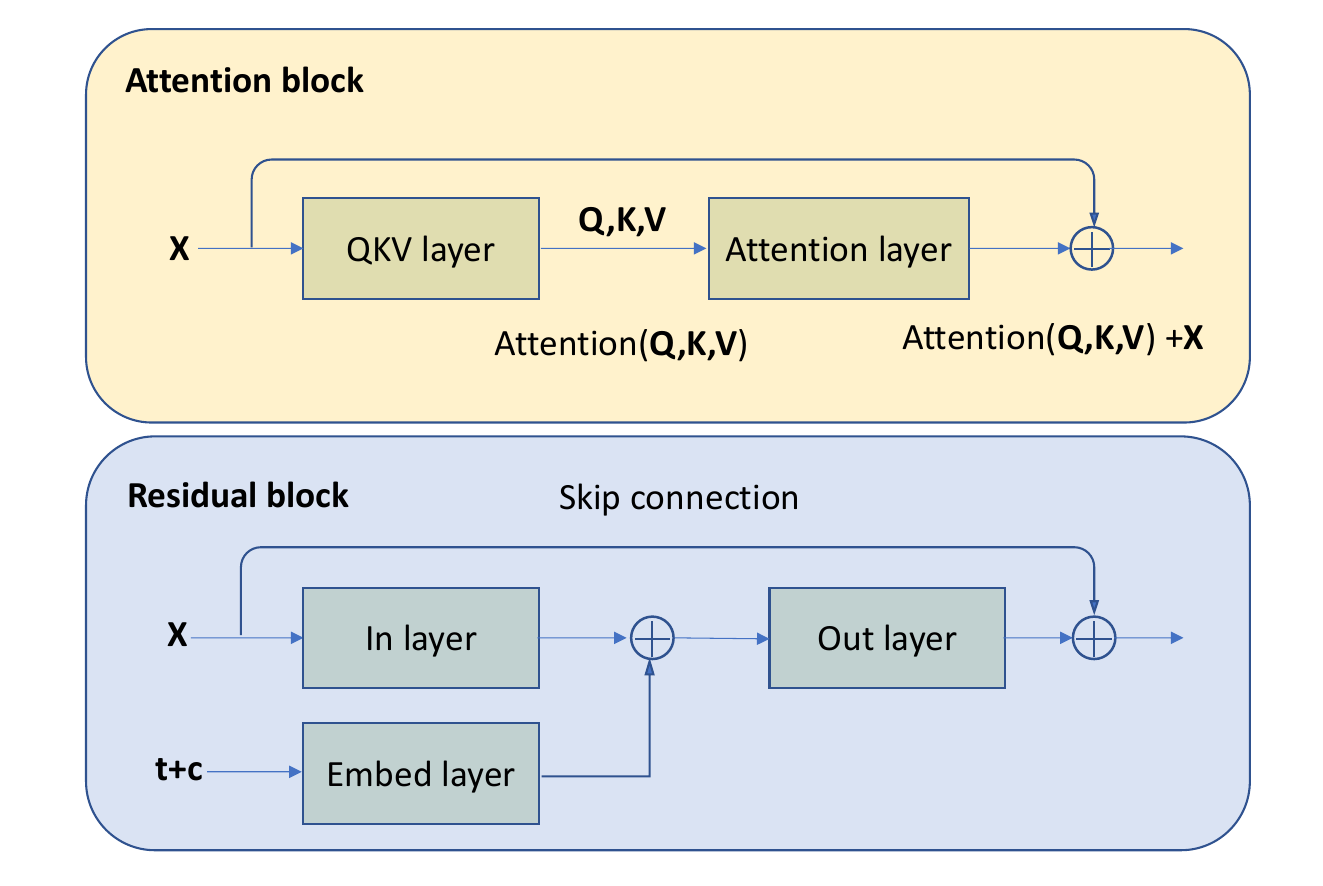}
 \caption{The attention block and residual block in the U-Net. Attention block: the 1D convolutional QKV layer takes the group normalization \citep{Wu2018GroupN} of the previous input $x$ and outputs the query $Q$, keys $K$, and values $V$ which are used for the attention calculation. Residual blocks: it takes the linear embedding of the time $t$ and parameters $\mathbf{c}$ as the additional input, where the time and parameter information has already been encoded by the sinusoidal and linear embedding, respectively. The skip connection here is an identity layer or a convolutional layer depending on whether the input and output channels are the same or not. The in layer contains the group normalization and 2D convolution, while the out layer contains the group normalization and SiLU \citep{2017arXiv170203118E} function. 
}
 \label{fig: block}
\end{figure*}

Our U-Net architecture mostly follows \citet{NEURIPS2021_49ad23d1}. It is a stack of residual blocks and downsampling layers, followed by a stack of residual blocks and upsampling layers. The downsampling is by a 2D convolutional layer with stride 2, while the upsampling is by a 2-times nearest interpolation and a following convolution layer. The skip connections are applied between the intermediate outputs with the same spatial size during the encoding and decoding processes, where the two outputs are simply concatenated together along the channel dimension. In addition, multi-head attention blocks with specific resolutions are also applied. In Fig.~\ref{fig: block}, we show the main structures of the residual block and attention block. The attention block applies a self-attention scheme, allowing spatial positions to attend to each other. This is done by the QKV attention layer \citep{NIPS2017_3f5ee243} which tends to find keys $K$ that match the query $Q$ and get the values $V$:
\begin{equation}
\operatorname{Attention}(Q, K, V)=\underset{s e q}{\operatorname{softmax}}\left(\frac{Q K^{\top}}{\sqrt{d_{c}}}\right) V,
\end{equation}
where $d_{c}$ is the channel size. The softmax function is along the sequence (flattened features) dimension of the key. The multi-head attention creates multiple (number of heads) sets of $QKV$ matrics and concatenates over the head dimension in the final step of the attention calculation. 

In this work, we apply 4 downsampling and upsampling layers so that the outputs have the resolution (spatial size) of $(64,64)$, $(32,32)$, $(16,16)$, and $(8,8)$, respectively. We use 3 residual blocks per resolution during the encoding process and 4 residual blocks per resolution during the decoding process. There are 96 base channels for the residual block. In each resolution the residual block has 1-, 2-, 3-, and 4-time base channels, except the output of the upsampling and downsampling layers which has the same channel as the input. We also apply the attention blocks with 4 heads at the resolution of $(16,16)$ and $(8,8)$. We use the initial learning rate $lr=1\times 10^{-4}$ which linearly decays with the rate $-lr/I_{tot}$, where $I_{tot}=600$ is the total training iterations. We also use the exponential moving average (EMA), $\bar \theta_t=\beta \bar \theta_{t-1}+(1-\beta) \theta_t$ for the stability of the model's convergence, where $t>1$ is the number of network parameter updates, $\bar \theta_{0}=\theta_0$ is initialized model parameters, $\bar \theta_t$ is the EMA of the previous updates, $\theta_t$ is the model parameters in the current update, and $\beta=0.995$ is the EMA rate. For our small number of training iterations, in order to make the most use of the EMA, we choose the ratio of the batch size to the training data size to be $1/20$, i.e., using the batch size of 8, 16, 32, 64, and 128 for the training sample size of 1600, 3200, 6400, 12800, and 25600, respectively. During training, we monitor the FSD based on a particular testing set $(0.200,0.800)$ which has the most distinguishable features in the images. Specifically, we generate 64 images with DDPM sampling at certain epochs during training and compare these generated images with 64 simulated ones with the FSD. We choose the models that have the smallest FSD at that point. Note that because this small number of images does not fully represent the real distribution, the FSD should have some fluctuations. So we also consider models that have FSD around this smallest value. Finally, We choose the checkpoint at the iteration $I=560$, where the FSD calculated from the generated images is around the smallest one when training on 25600 samples. We then use the checkpoints at the same iteration $I=560$ for all experiments with different training sample sizes, for simplicity.

\subsection{StyleGAN2}
\label{sec: gan arch}

Section \ref{sec: stylegan} explicates that the generator is comprised of a mapping network $f$ and a synthesis network $g$. $f$ is composed of two multi-layer perceptrons (MLP), with $f_1$ being an eight-layer structure that maps a Gaussian random vector $\mathbf{z}$ to a vector of length 512: $f_1(\mathbf{z})$. The other MLP, $f_2$, is a two-layer structure that maps astrophysical parameter $\mathbf{c}$ to a 256-length vector $f_2(\mathbf{c})$. The final style vector $\mathbf{w}$ is formed by multiplying half of the components in the output vector of $f_1(\mathbf{z})$ by $f_2(\mathbf{c})$. The synthesis network $g$ starts from a fixed layer with a size of $(512,2,2)$ and performs two convolutions before a $2\times$ upsampling. At the end of each convolution, Gaussian noise of the same size is injected into the feature map. After five rounds of upsampling, the feature map size grows from $(512,2,2)$ to $(256,64,64)$, with channel reduction applied to save memory usage. Prior to each upsampling, an additional convolution layer converts the current feature map to a final image with the corresponding size. For example, the convolutional layer converts the $(512,2,2)$ feature map to a $(1,2,2)$ pre-final image before the first convolution. By upsampling all pre-final images to the final size and summing them together, the final output image with a size of $(1,64,64)$ is obtained.

The discriminator consists of one input layer, five ResNet blocks, and a two-layer MLP. In our configuration, each block contains two convolutional layers that are stacked together. The feature map is downsampled by a factor of 2 between each ResNet block. The spatial dimension of the feature map is reduced from $(64,64)$ to $(4,4)$ after 5 ResNet blocks, and the feature map is then flattened into a lengthy vector. The MLP has two inputs: the aforementioned vector and astrophysical parameters. Its output is a score, where 0 indicates real and 1 denotes fake.

For training our network, we choose the batch size of 32 and training iterations of 90000. We use the default EMA rate in StyleGAN2 which is about 0.998. Different from DDPM, we monitor the FSD at all five testing sets as explained in Section~\ref{sec: stop}, and with 128 other than 64 generated images for FSD calculation. The reason for this larger testing batch is that StyleGAN2 has a faster sampling speed than DDPM. So we can choose a relatively larger number of images without introducing extra computational costs. The overtraining of GAN is known to often damage the performance, so we monitor the FSD, instead of choosing the checkpoint at the same iteration in DDPM, for each experiment. We find the optimal checkpoints at the iteration $I=25000, 55000, 41000, 83000, 57000$ for training sample sizes of 1600, 3200, 6400, 12800, and 25600, respectively.

\bsp	
\label{lastpage}
\end{document}